\documentclass[12pt]{article}

\usepackage[utf8]{inputenc}
\usepackage[T1]{fontenc}
\usepackage{lmodern}
\usepackage{eufrak}
\usepackage{microtype}

\usepackage{floatrow}
\usepackage[bottom]{footmisc}
\usepackage{parskip}
\usepackage[a4paper, margin=2cm]{geometry}

\usepackage{xcolor}
\usepackage{tikz}
\definecolor{mygrey}{HTML}{888888}
\definecolor{myoran}{HTML}{E69F00}
\definecolor{myblue}{HTML}{0072B2}

\usepackage{amsmath}
\usepackage{empheq}
\usepackage{mathtools}
\DeclareMathOperator{\BetaDistribution}{Beta}
\DeclareMathOperator{\BinomialDistribution}{Binom}
\DeclareMathOperator{\UniformDistribution}{Unif}
\DeclareMathOperator{\MultinomialDistribution}{Multinomial}
\DeclareMathOperator{\PoissonDistribution}{Poisson}
\DeclareMathOperator{\Beta}{B}
\DeclareMathOperator{\GammaFunction}{\Gamma}

\usepackage{booktabs}
\usepackage{siunitx}

\usepackage{xcolor}
\usepackage{graphicx}

\usepackage[autostyle]{csquotes}
\usepackage[british]{babel}

\usepackage{hyperref}

\newcommand*{\q}[1]{\enquote{#1}}
\DeclareRobustCommand*{\pcircle   }[1]{\tikz[baseline, x=1ex, y=1ex] \path [fill=#1] (.5,.5) circle (.5) ;}
\DeclareRobustCommand*{\pdisk     }[1]{\tikz[baseline, x=1ex, y=1ex] \path [draw=#1, thick] (.5,.5) circle (.5) ;}
\DeclareRobustCommand*{\prectangle}[1]{\tikz[baseline, x=1ex, y=1ex] \path [fill=#1] (.05,.05) rectangle (.95, .95) ;}

\title  {A simple Bayesian model to estimate proportions and ratios from
         count data with a hierarchical error structure with an application to
         droplet digital PCR experiments}
   \author {Elyas Mouhou%
         \thanks{Laboratoire GBCM (EA7528),
               Conservatoire national des arts et métiers,
               HESAM Université,
               2, rue Conté, 75003 Paris France.}%
\and
Vincent Audigier\thanks{CEDRIC, Conservatoire national des arts et métiers,
               HESAM Université,
               2, rue Conté, 75003 Paris France.}%
\and
Josselin Noirel\footnotemark[1]~\thanks{Corresponding author.}
}

\begin{document}

\maketitle

\begin{abstract}
  Experimental designs with hierarchically-structured errors are pervasive in many biomedical areas; it is important to take into account this hierarchical architecture in order to account for the dispersion and make reliable inferences from the data. This paper addresses the question of estimating a proportion or a ratio from positive or negative count data akin to those generated by droplet digital polymerase chain reaction experiments when the number of biological or technical replicates is limited. We present and discuss a Bayesian framework, for which we provide and implement a Gibbs sampler in R and compare it to a random effect model.
\end{abstract}

\vspace*{\stretch{1}}

\tableofcontents

\vspace*{\stretch{1}}

\newpage

\section{Introduction}

Routine estimation and hypothesis testing have little relevance in the context of data having a hierarchical error structure. This situation is however pervasive in the biosciences, in clinical science and in biotechnology research, fields where data incorporate two (sometimes more\footnote{For instance, quantitative proteomics generates quantifications where dispersion arising at the biological level, at the experimental level and the technical level — peptide and protein — see for instance \cite{Ow2009,Schwacke2009}.}) hierarchically-different sources of uncertainty: biological variation and technical variation.
When a hierarchical error structure is present, Student's $t$ test and the like must be thrown out and special techniques must be used. Some of those situations can be handled using mixed effect models where the aim really consists in estimating the higher-level parameters. Astute, hierarchically-built modelling can help increase the robustness of the estimates and give more credence to the results.

\begin{figure}
    \includegraphics[width=.7\linewidth]{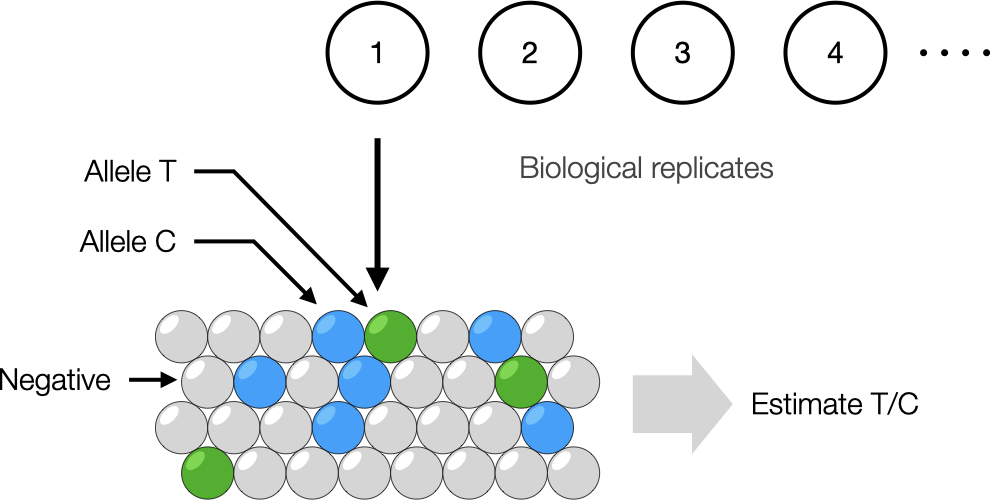}
    \caption{%
    ddPCR is a method of nucleic acid quantification that involves diluting then partitioning a sample into thousands of tiny droplets. Each droplet acts as a tiny reaction chamber and PCR amplification is carried out independently in each droplet.%
    %
    }
    \label{fig:ddPCR}
\end{figure}

Droplet digital polymerase chain reaction (ddPCR) is a technique that relies on dispersing DNA fragments into separate droplets for the amplification of isolated DNA molecules or, at the very least, a very limited number DNA molecules to take place. This allows researchers essentially to provide quantifications based on snapshot of near to single molecule resolution, be it for determining an allele frequency/allele ratio (often used in cancer research) or the fold-change in expression of some gene of interest (for targeted functional genomics). Fig.~\ref{fig:ddPCR} illustrates how ddPCR works. The idiosyncrasies of this experimental procedure are (1)~we're interested in a ratio hyperparameter and (2)~the observed data are count data with a latent Poissonian basis.

Let's illustrate this using a (not quite) hypothetical situation where $k$ biological replicates are run to estimate the allele ratio $r = A/B$ or the allele frequency $f = A/(A + B)$, for instance to gain an appreciation of how a tumour responds to a new treatment \cite{OlmedillasLopez2017}.
We expect the (unknown) allele frequency $p_i$ of each biological replicate to be dispersed around the population-wide value $f$.
The ddPCR data will register for each biological replicate~$i$:
(1)~$N_i$, the number of droplets where no amplification has taken place (negative droplets),
(2)~$A_i$, the number of droplets where allele $A$ only is amplified ($A$ positive),
(3)~$B_i$, the number of droplets where allele $B$ only is amplified ($B$ positive) and
(4)~$D_i$, the number of droplets where both alleles $A$ and $B$ are amplified (double positive).  Table~\ref{tab:example_ddPCR} provides an exemplar of ddPCR data.

In this paper, I describe a simple Bayesian approach to dealing with ddPCR data points and I provide a Gibbs sampler (packaged in an R library, see the Code availability section p.~\pageref{sec:Code_availability}) that will assist users in processing ddPCR data.

\begin{table}[ht]
    \sisetup{table-format=5.0}
    \begin{tabular}{@{} l *{5}{S} @{}}
    \toprule \hfill
	Droplets:   &{Positive C}&{Positive T}&{Double positive}&{Negative}&{Total}\\
    \midrule
    Replicate 1 &        186 &        166 &               1 &    14036 & 14389 \\
    Replicate 2 &        159 &        124 &               1 &    14773 & 15057 \\
    Replicate 3 &       6337 &       6054 &            2474 &      738 & 15603 \\
    \bottomrule
    \end{tabular}
    \caption{Example of ddPCR results.}
    \label{tab:example_ddPCR}
\end{table}

\section{Methods}


\subsection{Binomial approximation}

\begin{figure}
    \includegraphics[width=.3\linewidth]{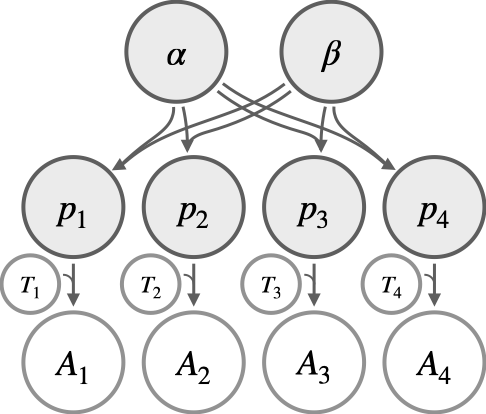}
    \caption{Simple hierarchical model used to perform Bayesian inference of ddPCR data.}
    \label{fig:model}
\end{figure}

Let's describe a hierarchical model for the ddPCR data, illustrated in Fig.~\ref{fig:model}.

Suppose that the fraction of $A$ alleles follows a beta distribution $B(\alpha, \beta)$, the two hyperparameters $\alpha$ and $\beta$ controlling both the central tendency and the dispersion across biological replicates. Ultimately, the parameters $\alpha$ and $\beta$ are to be estimated so as to provide an estimate of the allele ratio $\alpha/\beta$.  The fraction $p_i$ of $A$ positive droplets in biological replicate $i$ thus is drawn from distribution $\BetaDistribution(\alpha, \beta)$; the fraction of $B$ positive droplets is denoted $q_i = 1 - p_i$.

In order to facilitate the implementation of the model, (1)~we ignore the generally-negligible double positive droplets $N_i$, (2)~we assume that each positive droplet contains a unique DNA molecule and (3)~we assume that the total number of positive droplets $T_i = A_i + B_i$ is an observational constant. If follows that $A_i$ is simply drawn from a binomial distribution of parameters $\BinomialDistribution(p_i, T_i)$. Note these working hypotheses will be discussed towards the end of the paper (cf.\ Discussion, p.~\pageref{sec:Discussion}).

In order to fully specify the model, we include an improper, uninformative prior on $\alpha$ and $\beta$: both parameters are assumed to be uniformly distributed over $(1, +\infty)$. This ensures that proportions $p_i$'s are, in the worst case scenario, distributed uniformly over $(0, 1)$. This uninformative prior appears in practice to be both efficient and to provide a robust basis for the interpretation of a variety of situations:
\begin{itemize}
\item when the ratio is expected to be low because of low cellularity of a tumour for de~novo mutations,
\item when the ratio is expected to be close to $1$ because the somatic cells are heterozygous at a given locus exhibiting a single nucleotide polymorphism,
\item when the ratio can be pretty much anything in the context of differentiel expression measurements.
\end{itemize}

\subsection{Implementation of an approximate Gibbs sampler}

Our final model is as follows:
\begin{subequations}
\begin{align}
    \alpha &\sim \UniformDistribution(1, +\infty) \\
    \beta  &\sim \UniformDistribution(1, +\infty) \\
    p_i    &\sim \BetaDistribution(\alpha, \beta) \\
    A_i    &\sim \BinomialDistribution(p_i, T_i)
\end{align}
\end{subequations}
Implementing a Gibbs samples implies that we write the conditional probabilities:
\begin{subequations}
\begin{align}
    P(p_i    \mid \alpha, \beta) &\sim \BetaDistribution(\alpha + A_i, \beta + B_i) \\
    P(\alpha \mid \beta,  p_i)   &\propto \Beta(\alpha, \beta)^{-k} \times P^\alpha \label{eq:alpha} \\
    P(\beta  \mid \alpha, p_i)   &\propto \Beta(\alpha, \beta)^{-k} \times Q^\beta  \label{eq:beta}
\end{align}
\end{subequations}
where
$\Beta(a, b)$ is the beta function,
$K$ is the number of replicates,
$P$ is the product $P = \prod p_i$ ($1 \leq i \leq k$) and
$Q$ is the product $Q = \prod q_i$ ($1 \leq i \leq k$).
The main difficulty here is that, to the best of our knowledge, the probability density function that appears in \eqref{eq:alpha} and parametrised by $\beta$, $k$ and $P$ as follows ($C$ a normalisation factor)
\begin{equation}
P(\alpha; \beta, k, P) \mapsto C \times \Beta(\alpha, \beta)^{-k} \times P^\alpha \quad \alpha > 1
\end{equation}
isn't a well-known probability distribution.  (The same applies to \eqref{eq:beta}.)  In order to keep the sampler as simple as possible, we propose the following solution: the values of $\alpha$ are discretised and restricted to integers greater than or equal to $1$; additionally we place an upper bound, some integer $M$, which can be set manually or by using some numerical heuristics (see Appendix~\ref{sec:upper_bound}). In practice, upper bounds such as $M = 1000$ can be assumed and one can check in retrospect whether values sampled from the distribution are well below $M$.

\section{Application to ddPCR data}

\subsection{Allele ratio}

\begin{figure}
  \includegraphics[width=.43\linewidth]{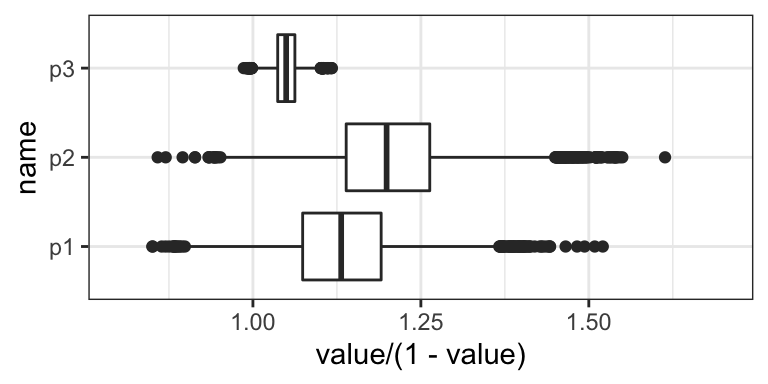}%
  \hspace*{.1\linewidth}%
  \includegraphics[width=.43\linewidth]{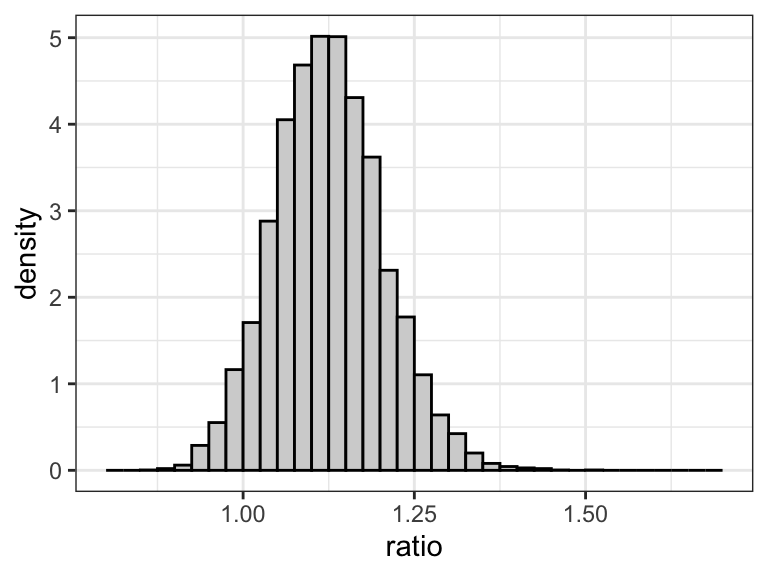}

  \medskip
  
  \makebox[.43\linewidth]{A: Boxplot of the $p_i$'s}%
  \hspace*{.1\linewidth}%
  \makebox[.43\linewidth]{B: Histogram of $\beta/\alpha$}
  
  \caption{Posterior distributions of each individual replicate's C/T ratio (as box~plots) and of the population-wide, idealised C/T ratio (as a histogram).}
  \label{fig:posteriors}
\end{figure}

In order to estimate the C/T given the data in Table~\ref{tab:example_ddPCR}, we ran the Gibbs sampler with the following parameters:
an upper~bound $M = 5000$,
$100$ burn-in cycles,
thinning by a factor $25$, so as to arrive at \num{10000} samples,
we can derive posterior distributions for $\alpha$, $\beta$ and each $p_i$.
One can focus on individual replicates to obtain credence intervals on the individual allele ratios $p_i/(1 - p_i)$ (cf.~Fig.~\ref{fig:posteriors}A) or on the broader picture, ie, the sampled values of the idealised allele ratio $\alpha/\beta$ (cf.~Fig.~\ref{fig:posteriors}B).

The advantage of a Bayesian framework here is that probability statements can be made on the allele ratio; for instance, we may calculate, assuming the general Bayesian hypotheses of the model broadly hold, the probability that the ratio is appreciably lower than some meaningful threshold. Let's consider the example provided in Table~\ref{tab:example_ddPCR} and analysed here: could the ratio C/T be conceivably lower than $1$? The answer given by the Gibbs sampler is that the probability $\Pr(r < 1 - \epsilon)$ is roughly 3\% (choosing somewhat arbitrarily the value $\epsilon = .01$).  This may paint a different picture from that given by a~naive Student $t$ test based on point estimates of the $p_i$'s (here this would give an unremarkable $p$-value of roughly $P = .09$).

\subsection{Measuring a low tumour fraction}
\label{sec:GNAQmut}

Droplet-based digital PCR is routinely used to estimate the tumour fraction based on a tumour-specific somatic mutation (\emph{GNAQ} Q209R).  The dataset below was intended to evaluate the technical sensibility of ddPCR in an experiment carried at the Institut Curie by diluting tumour DNA carrying the mutation with genomic DNA from healthy donors.

\begin{table}
    \sisetup{table-format=5.0}
    \begin{tabular}{@{} l *{5}{S} @{}}
    \toprule
	Droplets:   &{Positive C}&{Positive T}&{Double positive}&{Negative}&{Total}\\ \midrule
    Replicate 1 &          9 &       2245 &               1 &    10279 & 12534 \\
    Replicate 2 &          4 &       2414 &               0 &    10258 & 12676 \\
    Replicate 3 &          3 &       2102 &               1 &     9386 & 11492 \\
    Replicate 4 &          1 &       2449 &               1 &    11156 & 13607 \\
    Replicate 5 &          2 &       2739 &               2 &    12578 & 15321 \\
    Replicate 6 &          4 &       2442 &               1 &    12158 & 14605 \\
    \bottomrule
    \end{tabular}
    \caption{Example of ddPCR data with an extreme allele ratio $\mathrm{C} \ll \mathrm{T}$.}
    \label{tab:GNAQmut}
\end{table}

This example is interesting in that it illustrates a situation where there's an extreme imbalance between the two allele concentrations; as a consequence, point estimates are noisy and it appears crucial to use principled statistics to make robust inferences.

\begin{figure}
    \includegraphics[width=.7\linewidth]{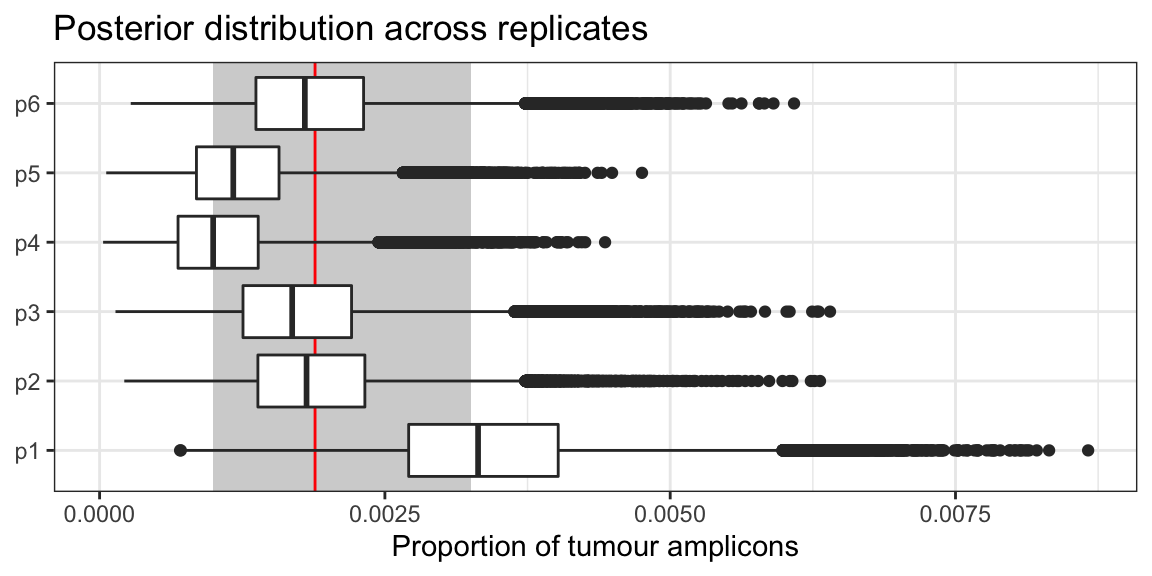}
    \caption{Posterior estimates for the $p_i$'s (as box~plots) and for the population-wide ratio (the shaded region represents 95\% credence interval and the red line corresponds to the point estimate).}
    \label{fig:GNAQmut}
\end{figure}

The model predicts for population-wide ratio a point estimate of $0.002$ a 95\% credence interval of $(0.001, 0.003)$.
Note in this example, it is worth checking that the upper bound~$M$ is well above the sampled values for $\alpha$ and $\beta$ ($M = 4000$ is satisfactory).

\subsection{Handling differential expression in cell lines}


If one is interested in the change in expression level of a gene (here, \emph{CLPTM1L}] caused by some allele introduced in a cell line (siNKX2.4), we have to measure the gene's expression against some reference gene's (here, \emph{GUSB1}).  The ddPCR technique is used to amplify \emph{CLPTM1L} or \emph{GUSB1} amplicons, instead of alleles.  The $\alpha/\beta$ ratio can be used to estimate the transcription level of \emph{CLPTM1L} relative to \emph{GUSB1}'s.  A change in transcription level therefore will be quantified as a ratio of ratios
\begin{equation}
  \frac { \{\alpha/\beta\}_{\mathmakebox[3pt][l]{\text{in siNKX2.4}}} }
        { \{\alpha/\beta\}_{\mathmakebox[3pt][l]{\text{in CTRL}}}       }
  \label{eq:ratio_diffexpr}
\end{equation}
The sorts of data we would have to deal with could look like those shown in Table~\ref{tab:CLPTM1L}.

\begin{table}
    \sisetup{table-format=5.0}
    
    \begin{tabular}{@{} l *{4}{S} @{}}
    \toprule \hfill
    &
    \multicolumn{2}{c @{}}{siCTRL cell line}   &
    \multicolumn{2}{c @{}}{siNKX2.4 cell line} \\
    \cmidrule(r{2mm}){2-3}
    \cmidrule(l{2mm}){4-5}
	Droplets:   &{CLPTM1L}&{GUSB1}&{CLPTM1L}&{GUSB1}\\
    \midrule
    Replicate 1 &    1905 &  841 &    2604 & 1063 \\
    Replicate 2 &    1640 &  742 &    1625 &  540 \\
    Replicate 3 &    1829 &  835 &    1550 &  600 \\
    Replicate 4 &    1581 &  811 &    1793 &  650 \\
    Replicate 5 &    3819 & 2139 &    4538 & 2243 \\
    \bottomrule
    \end{tabular}
    
    \caption{Example of differential gene expression data.}
    \label{tab:CLPTM1L}
\end{table}

\begin{figure}
    \includegraphics[width=.7\linewidth]{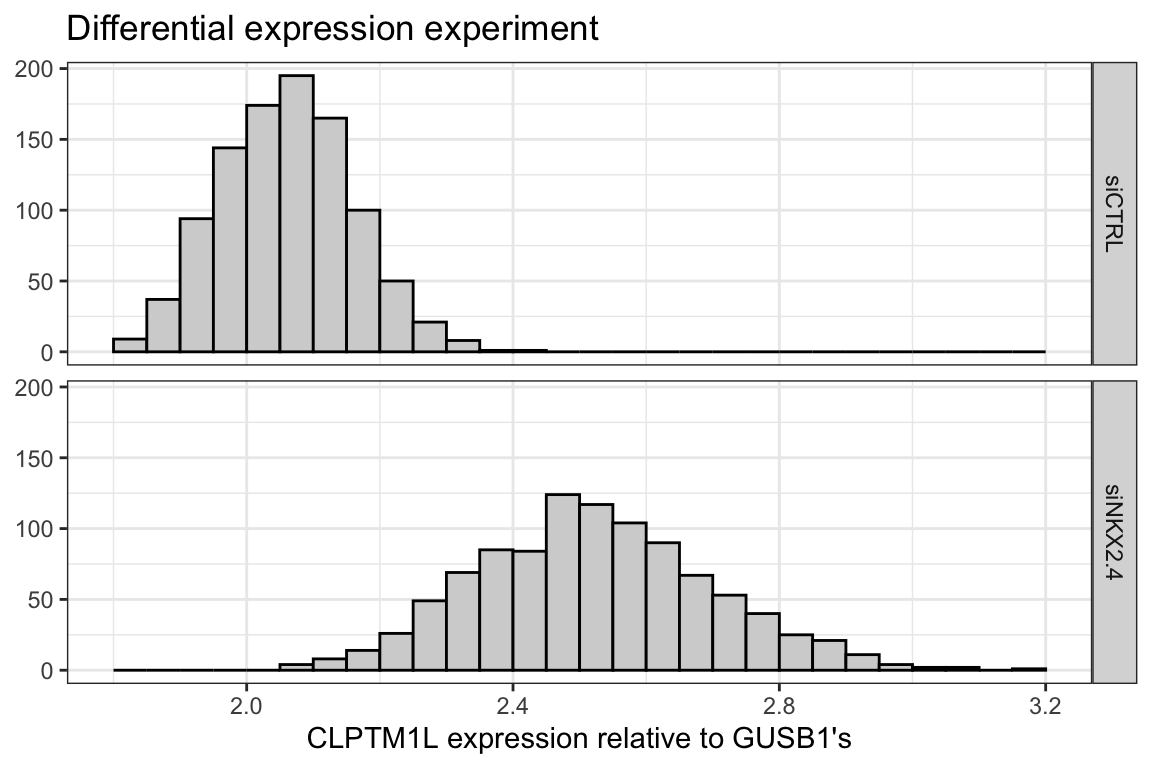}
    \caption{Example of differential gene expression data.}
    \label{fig:CLPTM1L}
\end{figure}

If we assume that data collected for individual cell lines are independent, we can simply evaluate the relative change in expression \eqref{eq:ratio_diffexpr} from samples obtained independently from both cell lines as per Fig.~\ref{fig:CLPTM1L}.  Samples are used to then calculate the relative change in expression from Eq.~\eqref{eq:ratio_diffexpr} and statistics can be obtained from them.  Applied to the specific Table~\ref{tab:CLPTM1L}, this gives a point estimate of $1.23$ for the relative increase in expression, 95\% interval: $(1.04, 1.45)$.

\section{Bayes appears more reliable than random-effects models}

Proportions with hierarchical structure can be modelled as via a logistic regression with each replicate-level proportion corresponding to a random effect.

We analysed one of Alan Agresti's datasets using our Bayesian model and compared its estimates to those of a random-effect model \cite{Agresti2012-fh}.

Table 13.2 (p. 499)
Chapter 13 \q{Clustered categorical data: random effects models}
Section 13.3 \q{Examples of random effects models for binary data}

\begin{figure}
    \includegraphics[trim=40 8 7 8, width=\linewidth]{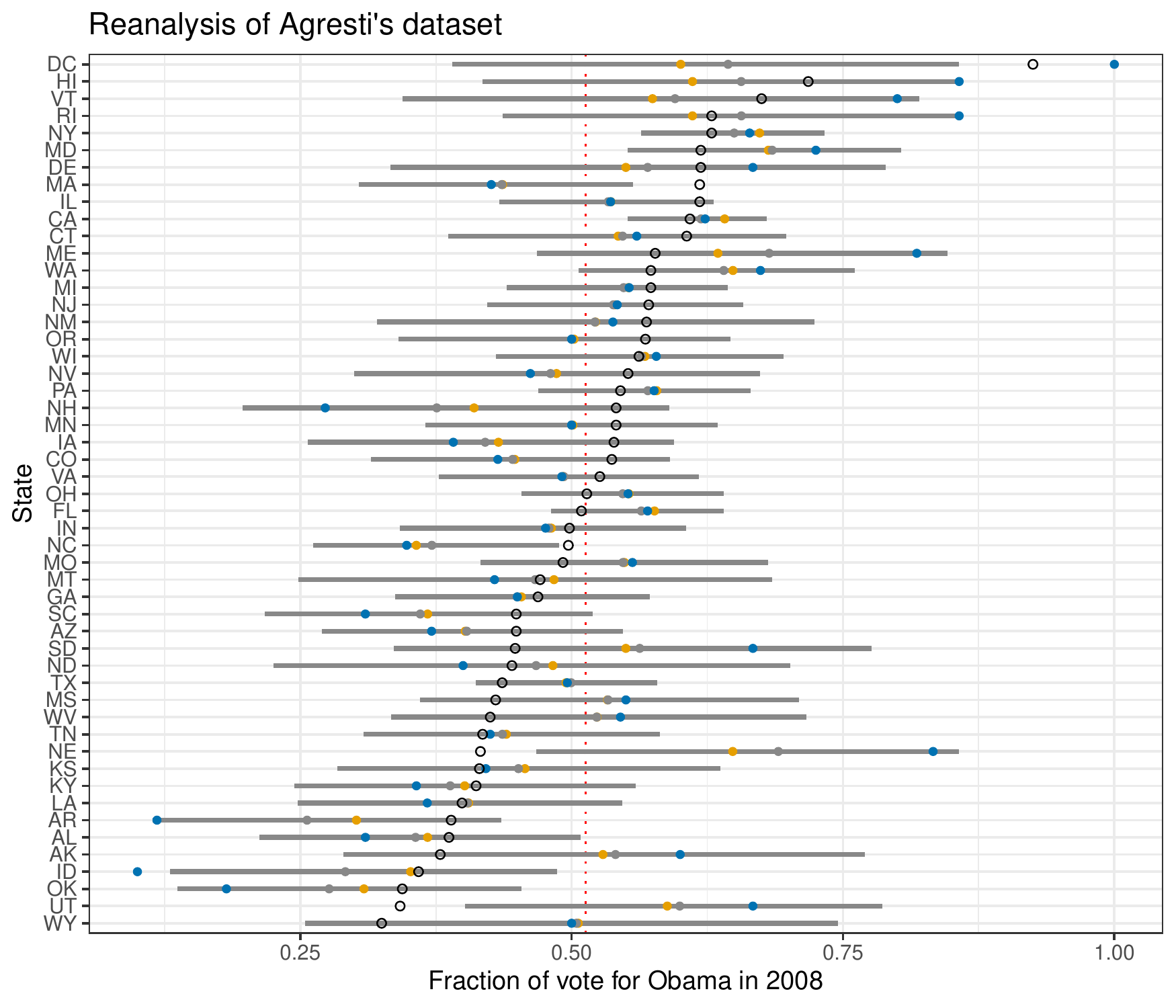}
    \caption{%
    Reanalysis of Agresti's dataset.
      \pcircle{mygrey}
      Grey discs and \prectangle{mygrey} grey bars:
      median and 95\% credence intervals of the state-level estimates.
      \pcircle{myoran}
      Orange discs: random-effect estimates.
      \pcircle{myblue}
      Blue discs: individual sample proportions.
      \pdisk{black}
      Black circles: ground truth values.
      \pcircle{red}
      Read: nationwise point estimate.}
    \label{fig:Agresti}
\end{figure}

This analysis consists in estimating the proportion of votes for Obama from state-level counts (the dataset is in fact made up).  The results are shown in Fig.~\ref{fig:Agresti}.
The bottom line here is that the random effect estimates agree with the Bayesian ones and the 95\% credence intervals are compatible the ground-truth value in all but five states.  The populationwide confidence and credence intervals are comparable:
$(0.466, 0.549)$ and
$(0.456, 0.548)$ respectively.

The Table~\ref{tab:example_ddPCR} gives another interesting test case since the number of biological replicates is low.
The Bayesian model described in this paper gives for $p_1$, $p_2$ and $p_3$ the following point estimates $0.531$, $0.546$ and $0.512$ respectively
and gives the following credence for the population-wide ratio $(0.495, 0.563)$.
In contrast, lme4 package fails to provide a robust estimation and reports a \q{boundary fit} and a \q{bad spline fit}: the 95\% confidence interval $(0.503, 0.547)$ is based on linear interpolation and appears extremely narrow given its upper end almost equal to one of the sample estimates (name $p_2$). The blme package converges but ends with an error message \q{Error in zetafun(np, ns) : profiling detected new, lower deviance}

\section{Discussion}
\label{sec:Discussion}

The model rests on few hypotheses and remains flexible enough to be of use in many situations.  In practice, it is only required to provide a safe upper~bound $M$ used to approximate the continuous distributions of \eqref{eq:alpha} and \eqref{eq:beta}. This facet of the model is further discussed in Appendix~\ref{sec:upper_bound} but researchers must bear in mind that, in the context of functional genomics experiments, expression data can be more strenuous on the model, since fold-changes can easily reach more extreme values.  Having said that, the default value ($M = 5000$) is meant to leave ample room for extreme fold-change situations; it appears sufficient for instance to handle the detection of the rare tumour amplicon of Table~\ref{tab:GNAQmut} detailed in section~\ref{sec:GNAQmut}. This was confirmed a~posteriori by plotting the posterior distribution of $\alpha$ and $\beta$ to see that they both are well below the $M$ upper bound.

When (1)~the allele ratio is very large or very small and, \emph{at the same time}, (2)~many droplets are positive, the binomial approximation (ie, assuming one amplicon by positive droplet) can become untenable and the estimation can significantly deviate from the ground truth.  Note that \emph{both} conditions are required for the model to break down; for instance, the fold-change obtained from analysing the \emph{GNAQ} mutation data (Table~\ref{tab:GNAQmut}) may be extreme, but the binomial approximation still holds.  Analysing the synthetic ddPCR data made of 50 replicates that comes with the bayesddpcr library, a dataset generated assuming an exact allele ratio of $6$ and where a sizeable fraction (32\% overall) of the droplets are positive, we overshoot the mark with a 95\% credence interval $(6.9, 7.6)$; including double positives in the count of A positive droplets and B positive droplets causes the model this time to underestimate the fold change, with a pretty narrow 95\% credence interval $(4.9, 5.2)$.  Note that a more complex model (see Appendix~\ref{sec:Poisson}) allows for a more satisfactory estimate to be obtained $(5.6, 6.2)$ (computationally this is about ten times more expensive to run).


In the context of measuring differential expression, assuming that the cell lines are independent allows one to steer clear of unnecessary complications and will be appear reasonable most of the time, although this means that batch effects cannot as such be accounted for.


\section{Code availability}
\label{sec:Code_availability}

The code is available as an R library from the author's GitHub repository
\url{https://www.github.com/JosselinNoirel/bayesddpcr/} (which can be installed through the devtools library); the code and datasets used to produce the figures is part of the repository, in the ‘article’ directory.

\section{Acknowledgments and author contributions}

We would lile to thank
Anne-Céline Derrien,
Caroline Hego,
Alexandre Houy,
Aurore Rampanou
Shufang Renault and Marc-Henri Stern for providing us with test cases and for stimulating discussions.

Elyas Mouhou developed the software, ran the analyses and wrote the article,
Vincent Audigier improved the model and the article,
Josselin Noirel  conceived the study, designed the model and wrote the article.

\appendix

\section{Upper bound from an approximation of the mode}
\label{sec:upper_bound}

The $M$ upper bound is set manually for simplicity's sake.  It must be noted however that a fairly accurate approximation can be obtained for the mode of the probability distribution of density function that appears in \eqref{eq:alpha} parametrised by $\beta$, $k$ and $P$ as follows ($C$ a normalisation factor):
\begin{equation}
P(\alpha; \beta, k, P) \propto \Beta(\alpha, \beta)^{-k} \times P^\alpha \quad \alpha > 1
\end{equation}
Using Stirling's approximation for the $\Gamma$ function ($\log \GammaFunction z \approx z \log z - z$), one can determine the mode occurs for a value $\alpha^*$ approximately given by
\[
    \alpha^{*} \approx \beta \cdot \frac{\sqrt[k]{p}}{\sqrt[k]{p} - 1}
\]
This value can be used as an initial value to iteratively seek an upper bound.

\section{A Poisson-based model specification}
\label{sec:Poisson}

The binomial model presented here can be improved to take into account the presence of multiple amplicons in a single droplet; denoting $\mathcal{N}_i$ the total number of droplets of the $i$-th replicate:
\begin{subequations}
\begin{align}
    \alpha         &\sim \UniformDistribution(1, +\infty) && \text{As above} \\
    \beta          &\sim \UniformDistribution(1, +\infty) \\
    p_i            &\sim \BetaDistribution(\alpha, \beta) \\[8pt]
    f_i            &\sim \UniformDistribution(0, 1) && \text{A dilution factor} \\[8pt]
    \mathcal{A}_i  &\sim \PoissonDistribution(f_i \cdot p_i) && \text{Number of A amplicons} \\
    \mathcal{B}_i  &\sim \PoissonDistribution(f_i \cdot q_i) && \text{Number of B amplicons}\ (q_i = 1 - p_i) \\[8pt]
    \mathfrak{n}_i &= \Pr(\mathcal{A}_i =    0 \land \mathcal{B}_i =    0) && \text{Negatives} \\
    \mathfrak{a}_i &= \Pr(\mathcal{A}_i \geq 1 \land \mathcal{B}_i =    0) && \text{`A' positives} \\
    \mathfrak{b}_i &= \Pr(\mathcal{A}_i =    0 \land \mathcal{B}_i \geq 1) && \text{`B' positives} \\
    \mathfrak{d}_i &= \Pr(\mathcal{A}_i \geq 1 \land \mathcal{B}_i \geq 1) && \text{Double positives} \\[8pt]
    N_i, A_i, B_i, D_i &\sim
    \MultinomialDistribution(\mathcal{N}_i,
                             \{\mathfrak{n}_i,
                               \mathfrak{a}_i,
                               \mathfrak{b}_i,
                               \mathfrak{d}_i\})
\end{align}
\end{subequations}
Note that a Gibbs sampler, albeit a more costly one, has been implemented in the bayesddpcr library.

\bibliography{refs}

\end{document}